\documentstyle[eqsecnum,aps]{revtex}
\begin{document}

\title{Charge exchange operators sum rules and proton-neutron T=0 and T=1 
pairing interactions}

\author{E. Moya de Guerra$^{1)}$, A.A. Raduta$^{2)}$, L. Zamick$^{3)}$, 
P. Sarriguren$^{1)}$}

\address{$^{1)}$Instituto de Estructura de la Materia,
Consejo Superior de Investigaciones Cient\'{\i }ficas,\\
Serrano 123, E-28006 Madrid, Spain \\
$^{2)}$Department of Theoretical Physics and Mathematics, 
Bucharest University, P.O.Box MG11, Romania and \\
Institute of Physics and 
Nuclear Engineering, Bucharest, P.O.Box MG6, Romania \\
$^{3)}$ Department of Physics and Astronomy, Rutgers University, 
Piscataway, New Jersey USA 08855-8019}

\maketitle

\begin{abstract}

We study sum rules for charge exchange operators of the Gamow-Teller 
and Fermi type and explore what can be learned from them about 
proton-neutron pairing in $T=0$ and $T=1$ channels. We consider both
energy weighted (EWSR) and non-energy weighted (NEWSR) sum rules. We
use a schematic Hamiltonian to stress the distinctive roles of $T=1$ 
and $T=0$ pair interactions and correlations in the sum rules.
Numerical results for $^{44}$Ti are presented for both schematic
and realistic interactions.
\end{abstract}

\section{Introduction}
\label{sec:level1}
The pairing interaction in finite fermion systems, as are atomic nuclei, has 
been widely investigated since the late fifties \cite{Bely,Kis}. Pairing 
interaction of alike nucleons has since then been taken into account in almost 
any microscopic calculation of various properties of medium and heavy nuclei. 
For a long time it was considered that proton-neutron pairing interaction may 
be ignored in medium and heavy nuclei due to the energy differences of proton 
and neutron Fermi levels. Despite this belief theoretical research in this 
field was performed \cite{Gos,Cam,God,Rad0}. Nowadays, at the era of heavy ion 
accelerators and of radioactive beams, allowing to study excitations of high 
spin states, as well as unstable nuclei close to the $N=Z$ line, the problem of 
proton-neutron ($\pi \nu $) pairing is being reconsidered and is a subject of 
much debate. 

For instance, in the case of $^{52}$Fe isotope the data about high spin states 
with $J\ge 10$ can be consistently explained if one assumes that from there on 
a new band with a broken proton-neutron pair starts \cite{Pov}. In particular, 
$\pi \nu $ pairing $T=0$  is considered to be an important source of high spin 
\cite{Good} by some groups. However, other groups \cite{Machi} claim that 
$\pi \nu $ pairing $T=0$  does not exist at all. 

Furthermore, $T=0$ pairing has as added interest that it is a new form of 
pairing, only possible in hadronic systems. It is clear that, for various 
reasons, the subject deserves attention and any contribution which might 
approach the answer concerning the existence of $\pi \nu $ pairing in $T=1$ 
and $T=0$ channels is welcome.

Here we study the echo of the $\pi \nu $ pairing of isospin $T=0$ and $T=1$, 
seen in two sum rules for $\beta$-decay operators. We consider both energy 
weighted (EWSR) and non-energy weighted (NEWSR) sum rules for these operators.

The sum rule approach is a useful classical tool to get global characterizations 
of the spectrum and response of a given nucleus to particular operators 
\cite{Boh}. The difference between  $\beta^-$ and $\beta^+$ non-energy weighted 
sum rules ($N-Z$ sum rule) was considered by several authors in connection with 
single $\beta$ decay and $(p,n)$ and $(n,p)$ reactions 
\cite{Ike1,Ike2,Auer,Zam,Ber,Gaar,Cha}. In ref. \cite{Zam} the charge exchange 
modes were studied on equal footing with the M1 mode, by comparing the results 
obtained with RPA and TDA approaches. Several approximations which might account 
for the missing strength of the Gamow-Teller (GT) states were discussed by 
Auerbach {\it et al.}, in ref. \cite{Auer}. Also effects of deformation 
\cite{Raddef,Sarr} on GT strengths have been studied within a selfconsistent 
microscopic framework.

In 1987, Cha considered the influence of the particle-particle channel of the 
two-body interaction on the $\beta^+$ strength \cite{Cha}. Indeed, he noticed 
a large sensitivity of the transition amplitude on the strength of the 
particle-particle interaction. This idea was extended by several groups to the 
double beta decay where all formalisms overestimated the transition amplitudes 
\cite{Fa}. To correct for this drawback new approaches going beyond QRPA have 
been proposed. While the QRPA approach obeys the $N-Z$ sum rule, any of the 
higher QRPA formalisms violates it. Thus, the first order boson expansion 
formalism \cite{Rad1} produces a sum rule which is by 20-25\% larger than $N-Z$. 
The renormalized QRPA \cite{Toi}, by contrary, yields a sum rule which is smaller 
than $N-Z$ by an amount of 20\%. In ref. \cite{Rad2} one of us (A.A.R.) 
formulated a first order boson expansion in terms of renormalized bosons and 
succeeded to bring the agreement of the sum rule with the $N-Z$ value within 10\%.

Here we would like to see in what direction are modified the $N-Z$ sum rules 
when the $\pi \nu $ pairing in a general sense is taken into account. More 
specifically, we address the question whether one can draw useful conclusions 
about the existence of $\pi \nu $ pairing from the EWSR and NEWSR of 
$\beta^-$ and $\beta^+$ strengths.

We accomplish this project according to the following plan. In Section 2, we 
treat the non-energy weighted sum rules and revisit Ikeda sum rule, while in 
Section 3 we focus on the energy weighted sum rule. In both cases the 
transition operators correspond to those of $\beta^-$ and $\beta^+$ transitions 
of Gamow-Teller and Fermi type. A numerical application for $^{44}$Ti is 
presented in Section 4. Final conclusions are summarized in Section 5.

\section{Non-energy weighted summed strengths and the Ikeda sum rule}
\label{secf:level2}

The total strength for the GT operator, acting in $\beta^-$ transitions 
and charge exchange $(p,n)$ reactions is defined as follows:

\begin{equation}
B_{GT}^{+} =\sum_{F}\left| \left\langle F\left| \vec{Y}_{\sigma }^{+}
\right| I\right\rangle \right| ^{2},
\label{def}
\end{equation}
where $|I\rangle$ denotes the ground state of the beta decaying nucleus, 
usually named as mother system, while $\{|F\rangle\}$ is a complete set of 
states describing the final nucleus, which can be reached with the GT dipole 
operator:

\begin{equation}
\vec{Y}_{\sigma }^{+} =\sum_{i}\vec{\sigma}_{i}t_{i}^{+}.
\end{equation}
Due to the completeness property of the set $\{|F\rangle\}$, the summation 
involved in Eq. (\ref{def}) can be performed and one arrives at the result:

\begin{equation}
B_{GT}^{+}=\left\langle I\left| \vec{Y}_{\sigma }^{-}\cdot 
\vec{Y}_{\sigma }^{+}\right| I\right\rangle .
\end{equation}
For the sake of completeness we specify the conventions we adopt for the 
isospin raising operator, $t^+$, and third component of the Pauli isospin 
matrix, $\tau_z$:

\begin{equation}
t^{+}\left| \nu\right\rangle =\left| \pi\right\rangle ;\quad t^{+}\left|
\pi\right\rangle =0; \quad
\tau _{z}\left| \pi\right\rangle =\left| \pi\right\rangle; \quad \tau _{z}
\left| \nu\right\rangle =-\left| \nu\right\rangle.
\end{equation}

The states $|\pi \rangle$ and $|\nu \rangle$ are single particle states 
specified by several quantum numbers describing the motion in coordinate
space, as well as spin and isospin degrees of freedom. The isospin quantum 
numbers are labeled by $\pi$ if the occupying nucleon is a proton and by $\nu$ 
if that is a neutron. Since the isospin operators do not change other quantum 
numbers but the isospin ones, we only specify the latter here. With these 
conventions it is clear that $B^+_{GT}$ describes the total strength for the 
$\beta^-$ decay mode.

Similarly, for the GT operator responsible for $\beta^+$-decay and $(n,p)$ 
reactions

\begin{equation}
\vec{Y}_{\sigma }^{-} =\sum_{i}\vec{\sigma}_{i}t_{i}^{-},
\end{equation}
the total strength is:

\begin{equation}
B_{GT}^{-}=\left\langle I\left| \vec{Y}
_{\sigma }^{+}\cdot \vec{Y}_{\sigma }^{-}\right| I\right\rangle. 
\end{equation}
Although, separately, the two total rates are not easy to be calculated, their 
difference can be exactly evaluated with a minimal effort:

\begin{equation}
B_{GT}^{+}-B_{GT}^{-} =\left\langle I\right| \sum_{i,j}\left( \vec{\sigma}_{i}
\cdot \vec{\sigma}_{j}\right) \left( t_{i}^{-}t_{j}^{+}-t_{i}^{+}t_{j}^{-}
\right) \left| I\right\rangle . 
\end{equation}
Using commutation relations for isospin operators

\begin{equation}
t_{i}^{-}t_{j}^{+}-t_{i}^{+}t_{j}^{-}=-\delta _{ij}\tau _{i}^{z},
\end{equation}
one obtains:

\begin{equation}
B_{GT}^{+}-B_{GT}^{-} =-\left\langle I\right| \sum_{i}\left( \vec{\sigma}_{i}
\right) ^{2} \tau _{i}^{z}\left| I\right\rangle =-6\left\langle I\right|
T_{z}\left| I\right\rangle =3\left( N-Z\right),
\end{equation}
where $N$ and $Z$ are the number of neutrons and protons, respectively,  
involved in the system under consideration. The equation obtained

\begin{equation}
B_{GT}^{+}-B_{GT}^{-} =3\left( N-Z\right),
\end{equation}
bears the name of its founder, we refer to it as Ikeda sum rule \cite{Ike1}. 
A similar sum rule holds also for Fermi transitions for which the transition 
operators are scalars against rotations in configuration space:
 
\begin{equation}
F^{\pm}=\sum_{i}t^{\pm}_i.
\label{211}
\end{equation}
  
Repeating the procedure used above for the GT transitions, one arrives 
at the expression:

\begin{equation}
B_{F}^{+}-B_{F}^{-}=\left( N-Z\right) .
\end{equation}

We should remark the fact that there is no approximation involved in deriving 
these Ikeda sum rules. This observation infers that these sum rules are very 
useful in testing the accuracy of any approximate scheme for the description 
of the mother nucleus ground state as well as of the ground and excited states 
in the daughter nucleus. They are also useful experimentally to define model 
independent quenching of GT strengths and serve to define the effective $g_A$ 
factor in nuclei.

In what follows we focus on how sum rules and energy weighted sum rules depend 
on pairing interactions. First of all, we notice that
$\vec{Y}_{\sigma }^{-}\cdot \vec{Y}_{\sigma }^{+}$ contains a one-body and a 
two-body operator:

\begin{equation}
\vec{Y}_{\sigma }^{-}\cdot \vec{Y}_{\sigma }^{+} =
\sum_{i}\vec{\sigma}_{i}^2\, t_{i}^{-}t_{i}^{+}+
\sum_{i\neq j}\vec{\sigma}_{i}\cdot \vec{\sigma}_{j}\, t_{i}^{-}t_{j}^{+}\, .
\end{equation}
The one-body term is simply three times the neutron number operator. This 
can be shown in coordinate representation

\begin{equation}
\sum_{i=1}^A\, 3t_{i}^{-}t_{i}^{+}=\sum_{i=1}^A\, 3\left( \frac{1}{2}-t_z^i
\right) =\frac{3}{2}\left( A-2T_z \right) =3N\, ,
\end{equation}
and can also be shown in second quantized notation as follows

\begin{equation}
\sum_{i=1}^A\, t_{i}^{-}t_{i}^{+}=\sum_{\alpha \rho , \alpha^\prime \rho^\prime}
\, \left\langle \alpha \rho \left| t^-t^+ \right| \alpha^\prime \rho^\prime 
\right\rangle a^+_{\alpha \rho} a_{\alpha^\prime \rho^\prime} = 
\sum_{\alpha \nu} a^+_{\alpha \nu} a_{\alpha^\prime \nu^\prime} = \hat{N}\, ,
\end{equation}
where $\rho$ stands for neutron $(\nu)$ or proton $(\pi)$ and $\alpha$ for
the remaining quantum numbers of the single-particle basis.

Then, we can write

\begin{equation}
B_{GT}^{+} =\left\langle I\right| \sum_{i,j}\vec{\sigma}_{i}\cdot 
\vec{\sigma}_{j}\, t_{i}^{-}t_{j}^{+}\left| I\right\rangle 
=3N+\left\langle I\right| \sum_{i\neq j}\vec{\sigma}_{i}\cdot 
\vec{\sigma}_{j}\, t_{i}^{-}t_{j}^{+}\left| I\right\rangle .
\label{eqbgtp}
\end{equation}
Note that the sum over $i,j$ runs over all nucleons, and the two-body term
can be of the same order as the one-body term and of opposite sign.

Likewise
\begin{equation}
\vec{Y}_{\sigma }^{+}\cdot \vec{Y}_{\sigma }^{-} =
\sum_{i}\, 3t_{i}^{+}t_{i}^{-}+
\sum_{i\neq j}\vec{\sigma}_{i}\cdot \vec{\sigma}_{j}\, t_{i}^{+}t_{j}^{-}=
3Z+\sum_{i\neq j}\vec{\sigma}_{i}\cdot \vec{\sigma}_{j}\, t_{i}^{-}t_{j}^{+}
\, ,
\end{equation}
and

\begin{equation}
B_{GT}^{-}=3Z+\left\langle I\right| \sum_{i\neq j}\vec{\sigma}_{i}\cdot 
\vec{\sigma}_{j}\, t_{i}^{-}t_{j}^{+}\left| I\right\rangle .
\label{eqbgtm}
\end{equation}
Therefore, $B_{GT}^{+}$ and $B_{GT}^{-}$ contain the same two-body term, which
does not contribute to  Ikeda sum rule, i.e., Ikeda sum rule cancels out all the 
two-body correlation effects in the ground state wave function. This is the 
reason why $B_{GT}^{\pm}$ depend on $\pi \nu $ pairing but the Ikeda sum rule 
does not.

The two-body term takes care of the correlations in the nuclear ground state,
including Pauli correlations. Its evaluation requires explicit knowledge of
the ground state. In particular we note that for a filled proton-neutron
$\ell-$shell the net contribution to $B_{GT}^{\pm}$ of the one-body plus
two-body terms must be zero. Thus, in shell model calculations Eqs.
(\ref{eqbgtp}) and (\ref{eqbgtm}) can be replaced by

\begin{equation}
B_{GT}^{+} = 3N_v+ \left\langle I \left| \sum_{i\neq j}\,\!^v \;\vec{\sigma}_{i}
\cdot \vec{\sigma}_{j}\, t_{i}^{-}t_{j}^{+}\right| I\right\rangle \; , 
\label{nv}
\end{equation}
\begin{equation}
B_{GT}^{-} = 3Z_v+ \left\langle I \left| \sum_{i\neq j}\,\!^v \; 
\vec{\sigma}_{i}\cdot \vec{\sigma}_{j}\, t_{i}^{-}t_{j}^{+}
\right| I\right\rangle \; , 
\label{zv}
\end{equation}
where $N_v$ and $Z_v$ denote valence nucleons outside closed $\ell-$shells
and the sums over $i,j$ run also over the valence nucleons only.
However, in calculations based on selfconsistent deformed mean field
like HFB and QRPA one has to consider the sums over all nucleons as in
Eqs. (\ref{eqbgtp}) and (\ref{eqbgtm}), as protons and neutrons occupy
slightly different orbitals and moreover with an underunity occupation
probability. We note that by definition $N_v-Z_v=N-Z$.

Similarly, for the Fermi total rate we obtain:

\begin{equation}
B_{F}^{+}=\left\langle I\right| \sum_{i,j}t_{i}^{-}t_{j}^{+}\left|
I\right\rangle =N+\left\langle I\right| \sum_{i\neq j}t_{i}^{-}t_{j}^{+}
\left| I\right\rangle ,
\end{equation}
and 

\begin{equation}
B_{F}^{-}=Z+\left\langle I\right| \sum_{i\neq j}t_{i}^{-}t_{j}^{+}\left| 
I\right\rangle ,
\label{eqbfm}
\end{equation}
where again the two-body term takes care of Pauli correlations. For the Fermi 
case we have that the sum of $\beta^+$ and $\beta^-$ strengths can also
be easily calculated as

\begin{equation}
B_{F}^{+}+B_{F}^{-}=\left\langle I\right| T^-T^+ + T^+T^-\left|
I\right\rangle = 2\left\langle I\right| \vec{T}^2 - T_z^2
\left| I\right\rangle .
\end{equation}
Thus, if we assume that the ground state has good isospin $T=|T_z|=|N-Z|/2$,
we end up with the sum rule
\begin{equation}
B^+_F + B^-_F =\left| N-Z \right| \, ,
\end{equation}
which together with the Ikeda sum rule, gives the well known results
\begin{equation}
B^+_F = N-Z \ ,\quad B^-_F =0\, ,
\end{equation}
assuming $N>Z$, in the SU(2) limit.

Thus, in the exact SU(2) limit, i.e., in the limit where isospin is an exact
symmetry, the two-body term contributes subtracting a quantity equal to the
maximum number of proton-neutron pairs in the A-body system. This is equal
to the Pauli correlation. Indeed the same result is obtained in the limit of
Slater determinantal wave function with identical proton neutron orbitals.
For each $\pi$ and $\nu$ occupying the same single particle state their total
isospin has to be zero due to antisymmetrization.
Because of Pauli principle the antisymmetrized product wave function contains
the maximum possible number ($Z$, for $N>Z$) of $\pi \nu$ pairs coupled to $T=0$,
and each pair contributes with $-1$ to the two-body term. Indeed it is easy to
show that for a $\pi \nu $ pair, the following identity holds:

\begin{equation}
t_{\pi }^{-}t_{\nu }^{+}=\vec{t}_{\pi }\cdot \vec{t}_{\nu }-t_{\pi
}^{z}t_{\nu }^{z}+i\left( \vec{t}_{\pi }\times \vec{t}_{\nu }\right) _{z} .
\end{equation}
Denoting by $\vec{T}_{\pi\nu}$ the $\pi \nu $ pair isospin and taking care of 
the fact that the nucleon isospin is $\frac{1}{2}$ one obtains:

\begin{equation}
t_{\pi }^{-}t_{\nu }^{+}
=\frac{1}{2}\left( \vec{T}_{\pi \nu }^{2}-\frac{3}{2}\right) +\frac{1}{4}
+i\left( \vec{t}_{\pi }\times \vec{t}_{\nu }\right) _{z}.
\label{tmtp}
\end{equation}
Now let us apply the operator $t_{\pi }^{-}t_{\nu }^{+}$, given by 
Eq. (\ref{tmtp}), on a $\pi \nu $ pair coupled to isospin $T$, 
$\left( \pi \otimes \nu \right) _{TT_z=0}$. The result is:

\begin{equation}
\left\langle \left( \pi \otimes\nu \right) _{T0}\right|
t_{\pi }^{-}t_{\nu }^{+}\left| \left( \pi \otimes \nu \right) _{T0}
\right\rangle = \left( \frac{1}{2}T\left( T+1\right) -\frac{1}{2}\right) 
=\frac{2T-1}{2},\, 
\end{equation}
valid for $T=0,1$.

Here we used the fact that the last term in Eq. (\ref{tmtp}) is antisymmetric 
and has zero expectation value in a state of given symmetry,

\begin{equation}
\left\langle \left( \pi \otimes \nu \right) _{T0} \right|
\left( \vec{t}_{\pi }\times\vec{t}_{\nu }\right)_z \left| \left( \pi 
\otimes \nu \right) _{T0}\right\rangle =0.
\end{equation}

An immediate consequence of this property is that the operator
$i\left( \vec{t}_{\pi }\times \vec{t}_{\nu }\right) _{z}$ does not contribute 
to the matrix element 
$\left\langle I\right| \sum_{\pi \nu }t_{\pi }^{-}t_{\nu }^{+}\left| I\right\rangle $
if the ground state has no isospin mixing. Hence, the operator 
$\sum_{i\neq j}t_{i}^{-}t_{j}^{+}$, acting on the ground state will give a factor 
$1$ for any $\pi \nu $ pair coupled to $T=1$ and a factor $-1$ 
for any $\pi \nu $ pair coupled to $T=0$. Thus, for Fermi like $\beta^{\pm}$ summed 
strength one obtains:

\begin{equation}
B_{F}^{+}=N-Z+ {\mathcal{C}}^{\left( \pi \nu \right) _{T}}\ ,\quad
B_{F}^{-}={\mathcal{C}}^{\left( \pi \nu \right) _{T}}\ ,
\label{eqbpf}
\end{equation}
where $\mathcal{C}^{\left( \pi \nu \right) _{T}}$ is a small correlation
function that takes into account small isospin mixing in the ground state
and that tends to increase with breaking of $T=0$ pairs.

 This is consistent 
with the fact that $T=1\; \pi \nu $ pairs are the only ones actually contributing 
to the states excited by the Fermi operator.

The evaluation of the two-body term for the GT strength

\begin{equation}
B_{GT}^{+}=3N+2\sum_{\pi \nu }\left\langle I\right| \vec{\sigma}_{\pi }
\cdot \vec{\sigma}_{\nu }\, t_{\pi }^{-}t_{\nu }^{+}\left| I\right\rangle .
\label{223}
\end{equation}
is more involved and requires exactly solvable models to be evaluated analytically.
Evaluation of $B^-_{GT}$ for valence nucleons in an $\ell-$shell in SU(4) and
SO(5) limits can be found in Ref. \cite{engel}.

In particular for the model discussed in Ref. \cite{engel} of $\mathcal{N}$ nucleon
pairs in an $\ell-$shell of degeneracy $2\Omega$, analytic expressions can
be given of the two-body matrix element. For the case of a standard isovector
spin-singlet hamiltonian the two-body matrix element in the ground state
with isospin $T=(N-Z)/2$ we get

\begin{equation}
\left\langle I\right| \sum_{i\neq j}\vec{\sigma}_{i}\cdot 
\vec{\sigma}_{j}\, t_{i}^{-}t_{j}^{+}\left| I\right\rangle =
-3Z\frac{(T+1)(\mathcal{N}+T+1)+\Omega +1/2}{(2T+3)(\Omega +1/2)}
\end{equation}

On the other hand, assuming a spatially symmetric correlated pair only, 
the contribution of the 
spin factor to the matrix element can be easily calculated and finally one 
gets:

\begin{eqnarray}
&&\left\langle (\pi\otimes\nu)_{T0}\right| \vec{\sigma}
_{\pi }\cdot \vec{\sigma}_{\nu }\, t_{\pi }^{-}t_{\nu }^{+}\left|
(\pi\otimes\nu)_{T0}\right\rangle =\left[ S(S+1)-\frac{3}{2}\right] 
\left[ T(T+1)-1\right] \nonumber \\
&& = -\frac{2T+1}{2}, \quad
{\rm for}\; T=0\;(S=1) \quad {\rm and} \quad T=1\; (S=0)\, .
\end{eqnarray}
Therefore, in a simple schematic model with independent $T=1$ and $T=0$ pairs, 
the total $B_{GT}^{+}$ strength will be
\begin{equation}
B_{GT}^{+}=3N-\left\{ 3{\mathcal{N}}^{\left( \pi \nu \right) _{T=1}}+
{\mathcal{N}}^{\left( \pi \nu \right) _{T=0}}\right\} .
\label{34_34}
\end{equation}

It is obvious that the $\beta^+$ strength can then be expressed as:
\begin{equation}
B_{GT}^{-}=3Z-\left\{ 3{\mathcal{N}}^{\left( \pi \nu \right) _{T=1}}+
{\mathcal{N}}^{\left( \pi \nu \right) _{T=0}}\right\} .
\label{226}
\end{equation}
Thus, for GT transitions, increasing both $T=1$ and $T=0$ number of 
$\pi \nu$ pairs ($\mathcal{N}$) 
leads to decreasing strengths. The important consequence of this is that 
using these expressions of total strengths for Gamow-Teller and Fermi 
transitions one can gain insight on the number of pairs with $T=1$ and $T=0$. 

We note that for fixed number of pairs and $2T=N-Z$, $B^-_{GT}$ decreases
as $T$ increases, in agreement with the behavior observed in Figs. 3 and 9
of Ref. \cite{engel}.

It is interesting to mention that these equations may also provide bounds 
for Gamow-Teller and Fermi transition total strengths. Since the total number 
of pairs in above equations cannot exceed the value of $Z$ (assuming $Z \le N$), 
from Eqs. (\ref{eqbpf})-(\ref{226}) one derives:

\begin{eqnarray}
0&\le & B^-_{GT}\le 3Z,\nonumber\\
3(N-Z)&\le & B^+_{GT}\le  3N,\nonumber\\
0&\le & \mathcal{C}^{(\pi \nu)_T}=B^-_F ,\nonumber\\
(N-Z) &\le & (N-Z)+ \mathcal{C}^{(\pi \nu)_T}=B^+_F ,
\label{bounds}
\end{eqnarray}
Comparing the second inequality of the second row in the above set of 
equations, with Eq. (\ref{223}) one concludes that the two-body term, which is 
washed out by Ikeda sum rule, has a negative contribution to $B^+_{GT}$. 
Moreover neglecting the two body term the $\beta ^-$ strength reaches its 
upper bound determined by the one body contribution. Thus, $T=0$ and $T=1$ 
$\pi \nu$ pairs tend to reduce the $B^{\pm}_{GT}$ strengths.
Owing to the previous discussion on Eqs. (\ref{nv},\ref{zv}), in shell model
calculations the upper bounds $3Z$ and $3N$ in Eqs. (\ref{34_34}) to 
(\ref{bounds}) can be
replaced by $3Z_v$ and $3N_v$, respectively.

\section{Energy weighted sum rules}
\label{sec:level3}

So far, no model Hamiltonian has been used, but we implicitly assumed that 
there are attractive proton-neutron interactions in $T=0\; (S=1)$ and  
$T=1\; (S=0)$ channels, favoring such pairs in the ground state. For our 
further consideration it is necessary to assume a many-body Hamiltonian $H$, 
including both one-body and two-body interactions. The one-body term is 
denoted by $H_0$ and the term describing the proton-neutron interactions 
will be denoted by $H_p$. Thus, we consider the following decomposition 
of $H$:

\begin{equation}
H=H_0+H_p.
\label{31}
\end{equation}
Moreover, we assume that $H_p$ is isoscalar and involves pairing in the 
two possible isospin channels $T=0$ and $T=1$, which may have different 
strengths $\omega_0$ and $\omega_1$. This will allow us to discriminate 
between the contributions brought by the two terms. For pragmatic reasons 
it is convenient to use schematic constant interactions where only the 
isospin dependence is explicit,

\begin{equation}
H_{p}=-\omega _{_{0}}\frac{1}{4}\left( 1-\vec{\tau}_{1}\cdot 
\vec{\tau}_{2}\right) -\omega _{_{1}}\frac{3}{4}\left( 1+\frac{1}{3}
\vec{\tau}_{1}\cdot \vec{\tau}_{2}\right) .
\end{equation}
Through direct calculations one proves that:

\begin{eqnarray}
H_p(\pi\otimes\nu )_{10}&=&-\omega_1 (\pi\otimes\nu )_{10},~~
H_p(\pi\otimes\pi )_{11}=-\omega_1 (\pi\otimes\pi )_{11},\nonumber\\
H_p(\nu\otimes\nu )_{1-1}&=&-\omega_1 (\nu\otimes\nu)_{1-1},
H_p(\pi\otimes\nu )_{00}=-\omega_0 
(\pi\otimes\nu )_{00}.
\end{eqnarray}
In other words, the first term of $H_p$ is a projector onto isospin $T=0$ 
pairs and the second term is a projector onto isospin $T=1$ pairs. If one 
wants to treat  the $T=1$ interaction between pairs of alike nucleons with 
distinct strength, one may supplement $H_p$ with an additional term:

\begin{equation}
H^{\prime}_p=-\frac{1}{2}\omega^{\prime}_1
\left( 1+ \tau ^{z}_{1}\tau ^{z}_{2}\right) ,
\end{equation}
which breaks isospin invariance.

Here we restrict our considerations to $H_p$ which does not discriminate between 
different channels of the nucleon nucleon pairing $T=1$ interaction. We assume 
that $T=0$ and $T=1$ pairing interactions are attractive and consequently 
$\omega _{0},\omega _{1}>0$. 

In what follows we shall study the energy weighted sum rules associated to the 
many body Hamiltonian described above and $\beta^-$ transitions. For Fermi 
transition, EWSR is determined by the transition operator F defined by 
Eq. (\ref{211}): 

\begin{eqnarray}
S^{F^{+}}&\equiv &\left( EWSR\right) _{F^{+}}\equiv
\sum_{F}(E_F-E_I)|\langle F|F^+|I\rangle |^2 \nonumber \\
&=&\frac{1}{2}\left\langle I\right| \left[
F^{-},\left[ H,F^{+}\right] \right] \left| I\right\rangle
\label{35}
\end{eqnarray}
The structure of the many body Hamiltonian given by Eq. (\ref{31}) induces 
the following split of the energy weighted sum rule:

\begin{equation}
\left( EWSR \right) _{F^{+}}=S_{0}^{F^{+}}+S_{p}^{F^{+}} .
\end{equation}
where the index $k(=0,p)$ indicates that the term corresponds to the 
Hamiltonian $H_k$.

The only terms which survive in the first commutator of Eq. (\ref{35}) 
for $H=H_p$ are:

\begin{eqnarray}
&&\left[ \vec{\tau}_{_{1}}\cdot \vec{\tau}_{_{2}},\sum_{j}t_{j}^{+}\right]
=\sum_{\lambda =x,y,z}\left\{ \tau _{_{1}}^{\lambda }\left[ 
\tau_{_{2}}^{\lambda },\sum_{j}t_{j}^{+}\right] +\left[ \tau _{_{1}}^{\lambda},
\sum_{j}t_{j}^{+}\right] \tau _{_{2}}^{\lambda }\right\} \nonumber \\
&=&\sum_{j}\sum_{\lambda =x,y,z}\left\{ \tau _{_{1}}^{\lambda}
\left( i\epsilon _{_{\lambda 2k}}i\tau _{2}^{k}-i\epsilon _{_{1\lambda k}}
\tau _{2}^{k}\right) \delta _{_{j,2}}+\left( 1\leftrightarrow 2\right)
\right\} =0.
\end{eqnarray}
where the standard notation for the antisymmetric unity tensor 
$\epsilon_{ijk}$, is used. Due to this relation it is manifest that: 

\begin{equation}
\left[ H_{p},F^{+}\right] =0.
\end{equation}
Therefore EWSR for the $\beta^-$ transition of Fermi type, is unchanged by  
the $\pi \nu $ pairing interaction:

\begin{equation}
S_{p}^{F^{+}}=0,
\end{equation}
as it is obvious because the Fermi operator commutes with any isoscalar operator.
If $H_0$ is not isospin invariant (a common case when the mean field single
particle states for protons and neutrons are different from each other),
adding only an attractive $T=0$ pairing interaction to $H_0$ will increase the 
number of $\pi \nu$ pairs coupled to $T=0$. This decreases the total strength, 
i.e., $ B_{F^{+}}<N$ (see Eq. (\ref{eqbpf})), while the EWSR is left unchanged. 
In order that this picture is achieved, some of the strength is to be moved to 
the higher energy states. The opposite is true for only $T=1$ $\pi \nu$ pairing 
interaction, i.e., the total strength is increased and a shift of the transition 
strength toward the low energy states is expected. Applying the hermitian 
conjugate operation to Eq. (\ref{35}) one also obtains that the proton-neutron 
interaction does not affect the EWSR associated to the $\beta^+$ transition, 

\begin{equation}      
S_{p}^{F^{-}}=0 ,
\end{equation}
and similar conclusions concerning the total strength and the migration of the 
strength to lower or higher energy states, as for the case of $\beta^-$, hold.
We stress that no migration of the strength is to be expected when $H_0$ is
isospin invariant and the ground state has isospin $T=T_z=(N-Z)/2$ both
before and after adding the pairing interaction.

Consider now the case of GT transitions. For this case also the two terms of 
the many body Hamiltonian ({\ref{31}) determine the following decomposition 
of the energy weighted sum rule:

\begin{equation}
S^{GT^{\pm}}=S^{GT^{\pm}}_0+S^{GT^{\pm}}_p.
\end{equation}

The term generated by the pairing interaction has the expression:

\begin{equation}
S_{p}^{GT^{+}} =\frac{1}{2}\left\langle I\right| \left[ 
\vec{Y}_{\sigma }^{-},\left[ H_{p,}\vec{Y}_{\sigma }^{+}\right] 
\right] \left| I\right\rangle .
\label{42}
\end{equation}

The double commutator involved in the above equation can be straightforwardly 
calculated if one uses the intermediate results:

\begin{eqnarray}
\left[ \vec{\tau}_{_{1}}\cdot \vec{\tau}_{_{2}},\vec{Y}_{\sigma }^{+}\right]
&=&\left( \vec{\sigma}_{1}-\vec{\sigma}_{2}\right)\left( 
\tau_{_{2}}^{z}\tau _{_{1}}^{+}-\tau _{_{1}}^{z}\tau _{_{2}}^{+}\right).
\nonumber\\
\left[ \vec{Y}_{\sigma }^{-},\left[ \vec{\tau}_{_{1}}\cdot \vec{\tau}_{_{2}},
\vec{Y}_{\sigma }^{+}\right] \right] &=&-\left( \vec{\sigma}_{1}-\vec{
\sigma}_{2}\right)^2 \left (\vec{\tau}_{_{1}}\cdot \vec{\tau}_{_{2}}+
\tau^z_{_1}\tau^z_{_2}\right). 
\end{eqnarray}

To calculate the average of the double commutator we have to know how each of 
the two factors, depending on spin and isospin respectively, acts on a given  
$(\pi\nu)$ pair.

\begin{eqnarray}
\left\langle (\pi\otimes \nu)_{T0}\right| (\vec{\tau}_{_{\pi}}\cdot 
\vec{\tau}_{_{\nu}}+ \tau^z_{\pi}\tau^z_{\nu}) \left| (\pi\otimes \nu)_{T0}
\right\rangle &=&4(T-1), \; {\rm for}~T=0,1,\nonumber\\
\left\langle (\pi\otimes \pi)_{11}\right| (\vec{\tau}_{_1}\cdot \vec{\tau}_{_2}+
\tau^z_1\tau^z_2) \left| (\pi\otimes \pi)_{11} \right\rangle &=&2,\nonumber\\
\left\langle (\nu \otimes \nu)_{1-1}\right|
(\vec{\tau}_{_1}\cdot \vec{\tau}_{_2}+\tau^z_1\tau^z_2) \left| 
(\nu\otimes \nu)_{1-1}\right\rangle &=&2,\nonumber\\
\left\langle (1\otimes 2)_S \right| \left( \vec{\sigma}_{1}-
\vec{\sigma}_{2}\right )^{2}\left| (1\otimes 2)_S\right\rangle&=& (-8S+12),  
\; {\rm for} ~S=0,1.
\end{eqnarray}  

Hence, for spatially symmetric correlated pairs (i.e., $T=0,S=1$ and $T=1,S=0$) 
we have that in a simple schematic model with independent $T=1$ and $T=0$ pairs

\begin{equation}
S_{p}^{GT^{+}}=\left (\omega _{0}-\omega _{1}\right )\left (2{\mathcal{N}}^
{\left ( \pi \nu \right )_{T=0}}-3\left({\mathcal{N}}^{\left( \pi \pi 
\right )_{11}}+{\mathcal{N}}^{\left ( \nu \nu \right )_{1-1}}\right )\right ).
\label{w01}
\end{equation}
This equation indicates that EWSR is influenced by both types of paring 
interactions, provided that the inequalities

\begin{equation}
\omega_0\ne \omega_1~~{\rm and} \quad {\mathcal{N}}^{\left( \pi \nu 
\right) _{T=0}}\ne \frac{3}{2}\left({\mathcal{N}}^{\left( \pi 
\pi \right)_{11}}+ {\mathcal{N}}^{\left( \nu \nu \right)_{1-1}}\right),
\end{equation}
hold.

If in Eq. (\ref{42}) the places of operators $\vec{Y}^+_{\sigma}$ and 
$\vec{Y}^-_{\sigma}$ are inter-changed, then one obtains the EWSR for the 
$\beta ^+$ transition. It is easy to show that for the $ \beta ^+ $ case, 
one has:
\begin{equation}
S_{p}^{GT^{-}}=S_{p}^{GT^{+}}.
\end{equation} 

Adding only an attractive $T=0$ force ($\omega_0 >0$) to the mean field 
Hamiltonian, $H_0$, will increase ${\mathcal{N}}^{(\pi\nu)_{T=0}}$. Therefore, 
the EWSR for GT operators will increase while the NEWSR for GT operators will 
decrease. The same is true when we add only an attractive interaction in $T=1$ 
channel. In this case it is interesting to note that the NEWSR decreases with 
increasing number of $T=1$ $\pi\nu$ pairs, while the EWSR does not depend 
directly on that number but increases with increasing number of proton-proton 
and neutron-neutron pairs. On the other hand when both $T=0$ and $T=1$ 
attractive pairing forces are present, their effects tend to cancel in the 
EWSR for Gamow-Teller.

\section{Application to $^{44}$Ti within $f_{7/2}$ shell}

Here we want to evaluate numerically the contribution of various interactions 
on the $GT$ strengths in a schematic single $j-$shell formalism.

In the single $j-$shell model for the $N=Z$ nucleus $^{44}_{22}$Ti  ($j=f_{7/2}$) 
the general form of a state with total angular momentum $I$, is:

\begin{equation}
\Psi ^{\alpha I}=\sum_{J_1,J_2}D^{\alpha I}(J_1J_2)
\left [(j^2)^{J_1}(j^2)^{J_2}\right ]^I,
\label{48}
\end{equation}
where $J_1$ is the total angular momentum of the two valence protons and $J_2$ 
of the two valence neutrons.

\subsection{Non-energy weighted strength}

For the transition from a given $I^\pi=0^+$ state to a 
given $I^\pi =1^+$ state, the GT strength is:
\begin{eqnarray}
B^+_{GT}(\alpha \rightarrow \beta )&=& 8\left| \langle j||\sigma ||j\rangle 
\right| ^2 \left| \sum_{J}
D^{\alpha 0}(J J)D^{\beta 1}(J J)\right. \nonumber \\
&&\times \left. \sqrt{2J+1}\left\{
 \matrix{1 & j & j \cr j &J & J} \right\} \right| ^2.
\label{bgtab}
\end{eqnarray}

If we sum over all final $1^+$ states, we get

\begin{eqnarray}
&&B^+_{GT}(\alpha )= (8-6\delta_{T,2})
\left| \langle j||\sigma ||j\rangle \right| ^2
\sum_{J} \left| D^{\alpha 0}(J J)\right| ^2 (2J+1)\left\{
 \matrix{1 & j & j \cr j &J & J} \right\} ^2 \nonumber \\
&=& 0.97959\left[D^{\alpha 0}(22)\right]^2+3.26531\left[D^{\alpha 0}(44)\right]^2
+6.85714\left[D^{\alpha 0}(66)\right]^2.
\label{expand}
\end{eqnarray}
Here, for $f_{7/2},\ | \langle j||\sigma ||j\rangle | ^2= 72/7$.

Note that in the single $j-$shell model space there are four $I^\pi =0^+$ states,
three with isospin $T=0$ and one with isospin $T=2$. There are three $I^\pi=1^+$
states, all with $T=1$. We will now calculate $B^+_{GT}$ for the $I^\pi =0^+$ 
ground state of $^{44}$Ti to $I^\pi =1^+\; T=1$ states in $^{44}$V. We consider 
several cases whose $B^+_{GT}$ values are given in Table 1.

{\it Case I. The isopairing Hamiltonian.}

We consider a simple isospin dependent Hamiltonian  of the form

\begin{equation}
H=-1+b\, \vec {t}_1.\vec {t}_2,
\label{hab}
\end{equation}
which is equivalent to the Hamiltonian considered in Section 3 but written
in the monopole form of French (see also Ref. \cite{caurier}).
When $b>0$, the state
$I^\pi =0^+\; T=2$ is pushed up in energy but still the three states 
$I^\pi =0^+\; T=0$ are degenerate. In particular, when $b=4$ we meet the 
situation of pure isopairing ($T=0$ pairing).

We again average over initial states

\begin{equation}
B^+_{GT}= \frac{1}{3} \sum_{\alpha (T=0)} \sum_{\beta} 
B^+_{GT}(\alpha (T=0)\rightarrow \beta)\, .
\end{equation}

In this case the total strength is obtained from the above equation by 
excluding from the summation over $\alpha$ the state 
$\alpha=|I^\pi=0^+,T=2\rangle$. In order to do that we need to know the 
coefficient $D^{T=2,J=0}(JJ)$, in the expansion (\ref{expand}). It turns out 
that these expansion coefficients are equal to the two particle fractional 
parentage:
\begin{equation}
\left|D^{T=2, J=0}(J,J)\right|=\left| \left ((j^2)J(j^2)J|\}j^40\right )\right|.
\end{equation}
$(D^{T=2}(0,0)=-0.5,\,D^{T=2}(2,2)=0.3127,\,D^{T=2}(4,4)=0.5,\,D^{T=2}(6,6)=0.6009)$. 
This is due to the fact that the $I^\pi=0^+, T=2$ state in $^{44}$Ti, can be 
obtained from the unique $I^\pi=0^+, T=2$ state in $^{44}$Ca by applying on it 
twice the raising isospin operator. Hence, one finds,
\begin{eqnarray}
B^+_{GT}&=&\frac{1}{3}\sum_{\alpha (T=0), \beta}B^+_{GT}(\alpha \rightarrow \beta )
= \frac{8}{3}\left| \langle j ||\sigma ||j\rangle \right| ^2 
\sum_{J}\left (1- \left ((j^2)J(j^2)J|\}j^40
\right) ^2  \right)\nonumber \\
&&\times (2J+1)\left \{
\matrix{1 & j & j  \cr j & J & J} \right \}^2 = 2.56\, .
\end{eqnarray}

{\it Case II. Same as Case I but with $b=0$.}

If we set $b=0$ then all four $J=0^+$ states are degenerate. To the total
strength in Case I we add the contributions of the $J=0^+\, T=2$ to all
final $J=1^+\, T=1$ states. The added value is

\begin{equation}
2\left| \langle j ||\sigma ||j\rangle \right| ^2 \sum_J (2J+1)
\left| \left( (j^2)J(j^2)J|\}j^40\right) \right|
(2J+1)\left\{
\matrix{1 & j & j  \cr j & J & J} \right\}^2 \, .
\end{equation}
When we average over the four initial states, we get $2.13$.

Now we consider another three cases obtained with pairing Hamiltonians which 
are different from the one given by Eq. (\ref{hab}).

{\it Case III. Standard $(J=0,\; T=1)$ pairing.} 

The case of isospin conserving $(J=0,\; T=1)$ pairing provides a non-degenerate 
ground state of $^{44}$Ti by means of a schematic many body Hamiltonian fixed 
by the condition

\begin{equation}
\langle (j^2)^{J}_T|V|(j^2)^{J}_T\rangle=-\delta_{J,0}\, \delta_{T,1}\; {\rm MeV}.
\label{flo}
\end{equation} 
This is just the reduced isospin model of Flowers \cite{flowers}. With condition
(\ref{flo}), the matrix elements in the space of two protons plus two neutrons, 
which is the case of $^{44}$Ti, become:
\begin{eqnarray}
H_{J_p,J^{\prime}_p}=-2\delta_{J_p,0}\delta_{J^{\prime}_p,0}-4(2J_p+1)
(2J^{\prime}_p+1)\left\{\matrix{j&j&J_p\cr j & j & J_p \cr 0 & 0 & 0 }\right\}
\left\{\matrix{j&j&J^{\prime}_p\cr j & j & J^{\prime}_p \cr 0 & 0 & 0 }\right\}.
\end{eqnarray}

Taking into account the expressions of the 9-j symbols with one vanishing row one 
arrives at the following matrix representation of the many body Hamiltonian.

\begin{equation}
H=-\frac{1}{16}\left[\matrix{33     & \sqrt{5} & 3          & \sqrt{13} \cr
                           \sqrt{5} &      5   & 3\sqrt{5}  &\sqrt{65}  \cr
                              3     &  3\sqrt{5}&  9        & 3\sqrt{13} \cr
                           \sqrt{13}& \sqrt{65} & 3\sqrt{13}& 13 }\right].
\end{equation}
This Hamiltonian has a non-degenerate ground state, one excited state at 0.75 MeV 
and two degenerate states at 2.25 MeV. The first eigenvalue is described by the 
state:

\begin{equation}
\Phi ^{0^+}=0.8660[00]^0+0.2152[22]^0+0.2887[44]^0+0.3469[66]^0.
\label{Phi}
\end{equation}
The total strength for $\beta^-$ transition from the non-degenerate ground state 
$0^+$ to any of the dipole states $1^+$ in the neighboring odd-odd nucleus
$^{44}V$, is given by Eq. (\ref{expand}) with $D^{\alpha 0}(JJ)$ given in 
Eq. (\ref{Phi}). The numerical value obtained can be seen in Table 1.

In the most often used case of like particle pairing and no $\pi \nu $ pairing

\begin{equation}
\left\langle (j^{2\nu})^{J}|V|(j^{2\nu})^{J}\right\rangle=
\left\langle (j^{2\pi})^{J}|V|(j^{2\pi})^{J}\right\rangle=
-\delta_{J,0}\; {\rm MeV},
\end{equation} 
which violates isospin conservation, one has that the coefficients in 
Eq. (\ref{48}) satisfy

\begin{equation}
D^0(J,J)=\delta_{J,0}
\end{equation} 
and therefore one sees from Eq. (\ref{expand}) that $B^+_{GT}=0$.
Though at first sight this result is surprising, the reason for that is that
we are introducing a pairing Hamiltonian in a single $j-$shell. Including 
only a single $j-$shell prevents pairing correlations to develop. So, we 
have to be careful in drawing too many conclusions from this particular result.

{\it Case IV. ($T=0,\; J=1$) pairing.}

In this case we consider a schematic many body Hamiltonian fixed by the condition

\begin{equation}
\langle (j^2)^{J}_T|V|(j^2)^{J}_T\rangle=-\delta_{J,1}\, \delta_{T,0}\; {\rm MeV}.
\end{equation} 

For the $J=0^+$ states of $^{44}$Ti, the matrix elements become:
\begin{equation}
H_{J_p,J^{\prime}_p}=-f\left( J_p\right) f\left( J^{\prime}_p\right)\, ,
\end{equation}
with
\begin{equation}
f\left( J_p\right) = 2\left\langle \left( j^2\right) ^{J_p}\left( j^2\right) ^{J_p}
\left| \right. \left[ \left( j^2\right) ^1 \left( j^2\right) ^1 \right]^0 
\right\rangle = 6 \left( 2J_p +1\right) 
\left\{\matrix{j&j&J_p\cr j & j & J_p \cr 1 & 1 & 0 }\right\} \, .
\end{equation}

Numerically, the Hamiltonian matrix for $J=0^+$ states in $^{44}$Ti for this
case is

\begin{equation}
H=\left[\matrix{ -0.18750 & -0.33940 & -0.20536 & +0.22535 \cr
                 -0.33940 & -0.61437 & -0.37173 & +0.40791 \cr
                 -0.20536 & -0.37173 & -0.22491 & +0.24681 \cr
                 +0.22535 & +0.40791 & +0.24681 & -0.27083 }\right].
\end{equation}

Proceeding as in the previous case, we find after diagonalization $B^+_{GT}=2.46$,
which is very close to the number obtained in Case I where $T=0$ also dominate.

The eigenfunctions and eigenvalues of this separable hamiltonian are easier to
obtain than for Case III $(J=0,\; T=1)$. The eigenvalue equation is

\begin{equation}
-f\left( J_p\right) \sum_{J_p^\prime} f\left( J_p^\prime \right)
D\left( J_p^\prime ,J_p^\prime \right) = \lambda D\left( J_p ,J_p \right) \, .
\end{equation}
If we multiply by $D(J_p,J_p)$ and sum, we get

\begin{equation}
-\left[ \sum_{J_p} D\left( J_p ,J_p \right) \right]^2 = \lambda \sum _{J_p}
D^2\left( J_p ,J_p \right) =\lambda \, .
\end{equation}
It is easy to show that 

\begin{equation}
D\left( J_p ,J_p \right) = \frac{f(J_p)}{\sqrt{-\lambda}} \, .
\end{equation}
From the properties of $9j$ and $6j$ symbols, we find

\begin{equation}
D\left( J_p ,J_p \right)= {\mathcal{N}}
\left( \matrix{63 \cr 5/\sqrt{5}  \cr 69 \cr -2/\sqrt{13} }\right) =
\left( \matrix{0.380 \cr 0.688  \cr 0.416 \cr -0.457 }\right) \, ,
\label{wf75}
\end{equation}
and then we get $B^+_{GT}=2.46$.

{\it Case V. Equal ($J=0,\, T=1$) and  ($J=1,\, T=0$) pairing}

\begin{equation}
\langle (j^2)^{J}_T|V|(j^2)^{J}_T\rangle=-\left( \delta_{J,1}\, 
\delta_{T,0} + \delta_{J,0}\, \delta_{T,1}\right) \; {\rm MeV}.
\end{equation} 

For the $J=0$ states of $^{44}$Ti the square matrix to be diagonalized is

\begin{equation}
H=\left[\matrix{ -2.250   & -0.47916 & -0.39286 & 0 \cr
                 -0.47916 & -0.92687 & -0.79100 & -0.09598 \cr
                 -0.39286 & -0.79100 & -0.78741 & -0.42923 \cr
                  0   & -0.09598 & -0.42923 & -1.08333 }\right].
\end{equation}
Note that the direct coupling $\langle [0,0]V[6,6]\rangle $ vanishes in
this case. For ($J=0,\, T=1$) pairing this matrix elements is -0.22535
whilst for ($J=1,\, T=0$) pairing it is +0.22535.

The ground state wave function is

\begin{equation}
\Psi=0.8256 [0,0] + 0.4047 [2,2] + 0.3726 [4,4] + 0.1259 [6,6]\, .
\end{equation}
We shall see that this is much closer to the realistic wave function
to be discussed in the next section. The summed $B(GT)$ strength is in
this case 0.73. This is smaller than the values for pure ($J=0,\, T=1$)
pairing and for pure ($J=1,\, T=0$) pairing. There is clearly a correlation
of strength when both interactions are present and with the same sign.

{\it Case VI. MBZ} 

This case corresponds to the situation considered by McCullen {\it et al.} in 
\cite{McC}, where the ground state is also  no longer degenerate. 
The quoted paper considers a more realistic two-body interaction whose matrix 
elements were fixed so that the spectrum in $^{42}$Sc is reproduced. 
In MBZ one has the integrated effects of ($J=0,\; T=1$) and ($J=1,\; T=0$)
pairing, as well as a quadrupole-quadrupole interaction. For this 
interaction the non-degenerate ground state of $^{44}$Ti is:
\begin{equation}
\Psi ^{0^+}=0.7608[00]^0+0.6090[22]^0+0.2093[44]^0+0.0812[66]^0.
\label{psi0}
\end{equation}
In the daughter odd-odd nucleus, the lowest level energy is a doublet 
$I^\pi=1^+,7^+$ at 0.6 MeV and a triplet $I^\pi=2^+,3^+,5^+$ at 1.5 MeV.
The GT transition takes place from the ground state of $^{44}$Ti, which is 
$0^+$, to the lowest state in $^{44}V$, i.e. $1^+$, which has the structure:
\begin{equation} 
\Psi ^{1^+}=0.9156[22]^1+0.3914[44]^1+0.0927[66]^1.
\label{Psi}
\end{equation}
Eqs. (\ref{psi0}) and (\ref{Psi}) determine the expansion coefficients 
$D^{\alpha I}$, which are to be introduced in the expression (\ref{bgtab}) 
in order to calculate the $B^+_{GT}$ strength for the transition 
$0^+ \rightarrow 1^+$.

Comparing the expressions (\ref{Phi}) and (\ref{psi0}), one sees that the 
coefficient $D^0(66)$ yielded in the schematic model described above, is much 
larger than that obtained by McCullen {\it et al.} \cite{McC}. The reason is 
that in ref. \cite{McC}, the formalism involves implicitly the $Q.Q$ interaction 
whose effect is to  push up the state $6^+$ which results in diminishing the 
overlap coefficient $D^0(66)$.

Summarizing, Case I corresponds to the situation of isopairing.
Case II is associated to 
the situation of no interaction, i.e., all four $J=0^+$ states are degenerate. 
Case III is the case of isospin conserving standard pairing, or Flowers model 
\cite{flowers}. Case IV is the case of ($J=1,\; T=0$) pairing, 
Case V is the case of an equal ($J=0,\, T=1$) and  ($J=1,\, T=0$) pairing
and finally Case VI is the realistic calculation from ref. \cite{McC}.
The results for the $\beta^-$ strengths corresponding to the cases described 
above, are collected in Table 1.

We note that although the total strength for the 'no interaction' case
is larger than that of $T=0$ isopairing, the averaged values are ordered
differently. The reason is that the number of states for the first case (4)
is larger than for the second case (3). The ordering of the total strengths
given in the first column of Table 1, is consistent with the fact that the
$T=0$ pairing decreases the total strength.
In fact, increasing the number of $T=0$ pairs the blocking effect is increased 
and the occupation probability decreased, which results in a quenching effect
for the total strength.

A more drastic reduction is obtained with 
isospin conserving standard pairing (Case III). An even stronger reduction is 
finally obtained in the realistic case considered in Case VI, which is caused by 
the fact that $Q.Q$ interaction, implicitly involved in the case VI, favors 
approaching the limiting $SU(4)$ symmetry for which $B^+_{GT}$ would vanish.
 
Note that for all cases considered, $\beta^+$ and $\beta^-$ strengths are equal.
Hence, pairing interaction reduces both $\beta^+$ and $\beta^-$ strengths.
These results complete those of ref. \cite{Cha} showing that the two-body 
dipole interaction in the particle-particle channel suppresses part of the 
$\beta ^+$  GT transition strength. Indeed, here we point out that a severe 
compression is taking place also for $\beta^-$ transition. 

\subsection{Energy weighted strengths}

We now consider results for several interactions:

{\it Case a.   ($J=0,\, T=1$) pairing interaction}

\begin{equation}
\left\langle \left( f_{7/2}^2\right) ^{J}_T V \left( f_{7/2}^2\right) ^{J}_T
\right\rangle = - |E(0)| \delta_{J,0}\, \delta_{T,1}\, .
\end{equation} 
The energy shifts and $B(GT)'s$ for states in $^{44}$Ti are as follows

\begin{eqnarray}
E(0^+)=-2.25 |E(0)|,\qquad && \nonumber\\
E(1_1^+)=-0.75 |E(0)|,\qquad && B(GT)_{0^+\rightarrow 1^+_1}=1.1507 \nonumber\\
E(1_2^+)=0,\qquad && B(GT)_{0^+\rightarrow 1^+_2}=0 \nonumber\\
E(1_3^+)=0,\qquad && B(GT)_{0^+\rightarrow 1^+_3}=0 \nonumber \\
\end{eqnarray}
$EWS=1.7260 |E(0)|$.

Note that there is no $B(GT)$ strength to the non-collective states, i.e.,
no strength to the states that are not shifted from their unperturbed
position by the ($J=0^+,\, T=1$) pairing interaction.

{\it Case b. ($J=1,\, T=0$) pairing interaction}

\begin{equation}
\left\langle \left( f_{7/2}^2\right) ^{J}_T V \left( f_{7/2}^2\right) ^{J}_T
\right\rangle = - |E(1)| \delta_{J,1}\, \delta_{T,0}\, .
\end{equation} 

Now we have

\begin{eqnarray}
E(0^+)=-1.2976 |E(1)|,\qquad && \nonumber\\
E(1_1^+)=-1.0262 |E(1)|,\qquad && B(GT)_{0^+\rightarrow 1^+_1}=0.03127 \nonumber\\
E(1_2^+)=-0.3190 |E(1)|,\qquad && B(GT)_{0^+\rightarrow 1^+_2}=2.4435 \nonumber\\
E(1_3^+)=0,\qquad && B(GT)_{0^+\rightarrow 1^+_3}=0 \nonumber \\
\end{eqnarray}
$EWS=2.3996 |E(1)|$.

Note that here also there is no strength to the non-collective state.
In contrast to case a) ($J=0,\, T=1$) pairing, we now have most of the
strength going to the second $1^+$ state, not the first.
This conclusion is consistent with our statement in Section 3 concerning the
migration of the transition strength when $T=0$ and $T=1$ pairings are 
considered separately.

{\it Case c. Equal ($J=0,\, T=1$) and ($J=1,\, T=0$) pairing}

\begin{equation}
\left\langle \left( f_{7/2}^2\right) ^{J}_T V \left( f_{7/2}^2\right) ^{J}_T
\right\rangle = - |E(0,1)| \left( \delta_{J,0}\, \delta_{T,1}\, +
\delta_{J,1}\, \delta_{T,0}\right) .
\end{equation} 
In this case

\begin{eqnarray}
E(0^+)=-2.6622 |E(0,1)|,\qquad && \nonumber\\
E(1_1^+)=-1.7131 |E(0,1)|,\qquad && B(GT)_{0^+\rightarrow 1^+_1}=0.604 \nonumber\\
E(1_2^+)=-0.3826 |E(0,1)|,\qquad && B(GT)_{0^+\rightarrow 1^+_2}=0.122 \nonumber\\
E(1_3^+)=0,\qquad && B(GT)_{0^+\rightarrow 1^+_3}=0 \nonumber \\
\end{eqnarray}
$EWS=0.8523 |E(0,1)|$.

This result confirms the statement made at the end of Section 3 saying that 
the $T=0$ and $T=1$ pairings interfere destructively for EWS.

According to Eq. (\ref{w01}), $S_{p}^{GT^{+}}=0$ for $\omega _{0}=\omega _{1}$.
However, here the value of EWS is small but not zero. The reason is that this
case is different from the one considered in Section 3.

{\it Case d. MBZ}

\begin{eqnarray}
E(1_1^+)=5.8146\, {\rm MeV},\qquad && B(GT)_{0^+\rightarrow 1^+_1}=0.5180 \nonumber\\
E(1_2^+)=8.5438\, {\rm MeV},\qquad && B(GT)_{0^+\rightarrow 1^+_2}=0.0337 \nonumber\\
E(1_3^+)=11.1393\, {\rm MeV}, \qquad && B(GT)_{0^+\rightarrow 1^+_3}\sim 0 \nonumber \\
\end{eqnarray}
$EWS=3.2995$ MeV.

Note that results for the overall $B(GT)$ strength for MBZ as well as the
($J=0^+,\, T=0$) $^{44}$Ti ground state wave function are much closer to the
results of the case of combined equal ($J=0^+,\, T=1$) and ($J=1^+,\, T=0$) pairing
than they are to the individual pairings. Hence, one needs both types of pairing
present in order to get realistic results.

Note in particular that for ($J=1^+,\, T=0$) pairing the [6,6] component of the
($J=0,\, T=0$) ground state wave function has the opposite phase of the same 
component in the more realistic MBZ wave function. The overlap
$<\Psi (MBZ), \Psi (J=1^+,\, T=0\,{\rm pairing})>$ is only 0.76, whilst
the overlap of MBZ with the mixed pairing case is 0.96.

\subsection{A note on the number of pairs}

For a system of $n$ nucleons in a single $j$ shell with total isospin $T$, 
we have the following
results for the number of pairs,

Total number of pairs: $n(n-1)/2$.

Number of pairs with isospin $T_0=0$: $n^2/8+n/4-T(T+1)/2$.

Number of pairs with isospin $T_0=1$: $3n^2/8-3n/4+T(T+1)/2$.

Hence, for the $T=0$ ground state of $^{44}$Ti ($n=4$) we have three $T_0=0$
pairs and three $T_0=1$ pairs.

Given a ground state wave function of the form

\begin{equation}
\Psi=\sum_{J_p} D(J_p,J_p) \left[ \left(j^2_{\pi}\right) ^{J_p} 
\left(j^2_{\nu}\right) ^{J_p} \right] ^{J=0}
\end{equation}
we can further obtain the number of pairs with total angular momentum $J_{12}$
$(J_{12}=0,1,2,3,4,5,6,7)$. The expression for the number of $J_{12}$ pairs is

\begin{equation}
2\left| D(J_{12},J_{12})\right| ^2 + \left| f(J_{12})\right| ^2
\end{equation}
where 

\begin{equation}
f(J_{12})=2(2J_{12}+1) \sum_{J_p} D(J_p,J_p) (2J_p+1) 
\left\{ \matrix{j & j & J_p \cr j & j & J_p \cr J_{12} & J_{12} & 0} \right\}
\end{equation}
Note that $D(J_{12},J_{12})$ is non-zero only for even $(J_{12},\, T_{12}=1)$
states.

We present results for the number of pairs with $(J_{12}=0,T_{12}=1)$,
$(J_{12}=1,T_{12}=0)$, and $(J_{12}=7,T_{12}=0)$
in Table 2 for various interactions. The total number
of pairs is of course 6, three with $T_{12}=0$ and three with $T_{12}=1$.

With the $(J=0,T=1)$ pairing interaction we get $2.25$ $(J_{12}=0,T_{12}=1)$ 
pairs. Recall that
the total number of $T_{12}=1$ pairs is three. The fact that this interaction
conserves isospin, forces the number of pairs to be less than three. There
are only $0.25$ $(J_{12}=1,T_{12}=0)$ pairs. The other $2.75$ are shared by
$J_{12}=3,5,7$.

With a $(J=1,T=0)$ pairing interaction the number of $(J_{12}=1,T_{12}=0)$
pairs increases considerably to $1.297$, whilst the number of $(J_{12}=0,T_{12}=1)$
pairs decreases to only $0.433$. When we have equal $(J=0,T=1)$ and $(J=1,T=0)$
pairing, we get $2.045$ and $0.618$ for the number of $(J_{12}=0,T_{12}=1)$
and $(J_{12}=1,T_{12}=0)$ pairs, respectively, result close to the MBZ result
of $1.736$ and $0.746$.

In general, there are more $(J_{12}=7,T_{12}=0)$ pairs than $(J_{12}=1,T_{12}=0)$
pairs. The number of $(J_{12}=7,T_{12}=0)$ is comparable to the number of
$(J_{12}=0,T_{12}=1)$ pairs.

It is interesting to note that for the $(J=1,\, T=0)$ pairing interaction, the
wave function given by Eq. (\ref{wf75}) can be written in another way,

\begin{equation}
\Psi = {\mathcal{N}} \left( 1- P_{p_1p_2}\right) \left( 1- P_{n_1n_2}\right)
\left[ \left( p_1n_1\right) _{J_{12}=1}\left( p_2n_2\right)_{J_{12}=1}
\right] ^{J=0} 
\end{equation}
where $P$ is the exchange operator. In other words, first couple one
neutron-proton pair to $J_{12}=1$ and then a second one. Couple the total
to angular momentum zero and then antisymmetrize. At first, this seems like
a very naive thing to do, but it works. Without the antisymmetry this looks like
two $J_{12}=1$ pairs, but the act of antisymmetrization complicates things
and gives us the result of Eq. (\ref{wf75}). We get only $1.297$ $J_{12}$ pairs.
It should be noted that for the $(J=0,\, T=1)$ pairing interaction one cannot get
a corresponding simple expression.


\section{The effect of the T=0 interaction on the Gamow-Teller
transitions in $^{44}$Ti}

Since the spin-flip configurations are important for Gamow-Teller transitions,
inclusion of the $f_{5/2}$ orbit is essential to discuss on Gamow-Teller
transitions as well as the Ikeda sum rule. In Sect. 4 the Ikeda sum rule is
satisfied due to the relations $B^+(GT) = B^-(GT)$ and $S_p^{GT^+}=S_p^{GT^-}$
in spite of the limitations of the configuration space.
Also without $f_{5/2}$, a condensate of $T=0$ pairs cannot appear \cite{engel}. 
In this section 
we discuss the effect of $T=0$ interaction including the full $fp$ shell.

We give results for $^{44}$Ti in which up to $t$ nucleons are excited from the
$f_{7/2}$ shell into higher shells. The $t=0$ case corresponds to a single $j$
shell calculation and $t=4$ to a full $fp$ shell calculation.

To find the effect of $T=0$ pairing we consider three interactions
\begin{itemize}
\item{I1.} The full FPD6 interaction \cite{richter}.
\item{I2.} Set the two particle matrix elements with isospin $T=0$ to zero
but keep the $T=1$ matrix elements of FPD6 unchanged.
\item{I3.} Set the $T=0$ matrix elements to -1.0 MeV but keep the $T=1$
matrix elements of FPD6 unchanged.
\end{itemize}

The idea is to see the effect of $T=0$ pairing by turning the $T=0$ interaction
off and on. The I3 interaction has nonzero $T=0$ matrix elements but does not
single out any particular ones. 

Just to be clear we distinguish between a constant
matrix element and a constant interaction. A constant interaction $a+bt_1\cdot t_2$
will have no effect on the wave functions or energy levels of a given total
isospin. It will cause shifts in energies of states of differing isospin.
A two-body matrix element can be written as

\begin{equation}
\left\langle \left[ j_1 j_2\right] ^{J}_T\ V\ \left[ j_3 j_4\right] ^{J}_T
\right\rangle
\end{equation} 
If we have a constant interaction then this will vanish if $(j_3,j_4)$ is not the
same as $(j_1,j_2)$.
What we are doing however in I3 is setting all $T=0$ two-body matrix elements
as above to -1.0 MeV. 

Those interactions have also been considered in Ref.
\cite{robinson}. The results are given in Table 3.
From Table 3 one can see that in the single $j$ shell calculation, the results
for I2 and I3 are identical. This can be understood by the fact that in this
small model space a constant matrix element and a constant interaction are
identical.

There is a large increase in $B(GT)$ when we go from $t=0$ to $t=1$. This is
because we can now have transitions from $f_{7/2}$ to $f_{5/2}$ (there were
not present in the $t=0$ calculation).

But the greatest interest lies in comparing the results for I1 and I2. When
the $T=0$ interaction is turned off (I2) the values of $B(GT)$ increase
considerably, especially in the full calculation ($t=4$). The change is from
1.236 to 3.316. Putting it the other way, reintroducing the $T=0$ matrix 
elements (I1) causes a large quenching of the summed $B(GT)$ strength. This
is also true in the single $j$ shell calculation but it is not as pronounced.
The pairing aspects of the $T=0$ interaction in FPD6 can be seen by examining
the spectrum of $^{42}$Sc where the excitation energies of the $J=1_1^+$ and
$J=7_1^+$ $T=0$ states are at 0.6 MeV but the $J=3_1^+$ and $J=5_1^+$ $T=0$
are at 1.5 MeV. We thus have the extreme spin states, lowest$-J$ and highest$-J$,
lower than the states with intermediate angular momentum.

To see that this separation is important we introduce the interaction I3 which
has constant $T=0$ matrix elements ($E=-1.0$ MeV). We see that the results
for I3 are similar to those of I2 -large values of $B(GT)$. Hence, the fine
details of the full interaction (I1), i.e., the pairing aspects for both low
and high spin, are important. They are needed to explain the strong quenching
of $B(GT)$.

\section{Conclusions}
In summary, we have investigated within a schematic model how $T=0$ and $T=1$
pairing correlations show up in the energy weighted and non-energy weighted sums
of $\beta^+$ and $\beta ^-$  strengths. We find that both EWSR and NEWSR depend 
differently on the number of $T=0$ and $T=1$ pairs in the ground state and a 
scheme is derived to assess the relative importance of $T=1$ and $T=0$ pairing
from experimental strength distributions obtained from $(n,p)$ and $(p,n)$ 
reactions. Formulas are given to get insight on pair numbers out of $\beta^-$ and 
$\beta^+$ strengths. In particular, we have found that attractive $T=0$ 
interactions alone tend to decrease the total GT strength, while increasing the 
EWS strength. Likewise $\pi \nu$ $T=1$ pairing alone.

The numerical results presented for schematic $j-$shell calculations in the 
$\beta^-$ GT decay of $^{44}$Ti show that indeed $T=0$ and $T=1$ pairing
correlations tend to decrease the GT strengths. The $f_{7/2}$ formalism is
restricted and the sum strengths are smaller than the total strengths obtained
in a major $N-$shell. Yet, this limited shell space allows to make simple
calculations that illustrate the effect of the various interactions.
The effect of $T=0$ pairing is best illustrated by the results shown on
calculations in the full $fp$ shell.

In the future it will be interesting to apply this scheme to analyze both
experimental data and realistic calculations. Calculations with more conventional
forms of the pairing interaction \cite{engel,poves}
are at present being considered.

To close this section we would like to say a few more words about previous 
results concerning the proton-neutron pairing and the single and double beta decay 
processes. As we mentioned, it was earlier noticed \cite{Cha} that the transition 
amplitudes for $\beta^+$ and double beta decay are strongly influenced by the two 
body interaction in the particle-particle channel. Analyzing the second 
quantization expressions for this interaction, we notice that it is nothing 
else but the ($T=0,\, J=1$) pairing interaction when one deals with Gamow-Teller 
transitions, and the ($T=1,\, J=0$) pairing interaction when  Fermi like transitions 
are considered \cite{Rad21}. We would like to stress that each of the processes 
mentioned above (Fermi and Gamow-Teller, $\beta^-$ and $\beta^+$) is influenced 
by both types of pairing. 

\vskip 1cm

\begin{center}
{\Large \bf Acknowledgments} 
\end{center}

This work was partially supported by MCyT (Spain) under contract numbers PB98-0676 
and BFM2002-03562, by  a U.S. Dept. of Energy Grant No. DE-FG02-95ER-40940 and by 
a NATO Linkage Grant PST 978158.

\vfill\eject

\vfill \eject

\begin{table}[h]

{\bf Table 1.} Summed $\beta^-$ strengths for $^{44}$Ti within $f_{7/2}$ shell with
various schematic interactions (see text).
\vskip .5cm

\begin{center}
\begin{tabular}{lcc}
\hline
Case & $B^+_{GT}$ & $B^+_{GT}$  \\
& Total & Averaged over \\
& & initial states \\
\hline
I. Isopairing                        & 7.674 &  2.56 \\
II. No interaction                   & 8.531 &  2.13 \\
III. Isospin conserving $(J=0,\, T=1)$ pairing  &  & 1.14 \\
IV. $(J=1,\, T=0)$ pairing           &  & 2.46 \\
V. Equal $(J=0,\, T=1)$ and $(J=1,\, T=0)$ pairing & & 0.73 \\
VI. MBZ                               &  & 0.55 \\     
\hline
\end{tabular}
\end{center}
\end{table}

\vskip 1cm

\begin{table}[h]

{\bf Table 2.} 
Number of pairs with $J_{12}=0,1$ and $7$ in a single $j$ shell 
calculation for $^{44}$Ti. Results are for interaction III: Isospin conserving 
$(J=0,\, T=1)$ pairing; IV: $(J=1,\, T=0)$ pairing; V: Equal $(J=0,\, T=1)$ and 
$(J=1,\, T=0)$ pairing; and VI: MBZ.

\vskip .5cm

\begin{center}
\begin{tabular}{cccc}
\hline
Interaction & \multicolumn{3}{c} {Number of $J_{12}$ pairs} \\
\cline{2-4}
&  $(J_{12}=0,\, T_{12}=1)$ &  
$(J_{12}=1,\, T_{12}=0)$ & $(J_{12}=7,\, T_{12}=0)$ \\
\hline
 III & 2.250 & 0.250 & 1.250 \\
 IV & 0.433  & 1.297 & 1.311 \\
 V & 2.045 & 0.618 & 1.654 \\
 VI & 1.736 & 0.746 & 1.948 \\
\hline
\end{tabular}
\end{center}
\end{table}

\vskip 1cm

\begin{table}[h]

{\bf Table 3.} The values of $B(GT)$ in $^{44}$Ti for the three interactions 
(I1,I2,I3) and up to $t$ nucleons excited from the $f_{7/2}$ shell.
\vskip .5cm

\begin{center}
\begin{tabular}{cccc}
\hline
$t$ & I1 & I2 & I3  \\
\hline
 0 & 0.627 & 0.840 & 0.840 \\
 1 & 3.192 & 4.109 & 3.495 \\
 2 & 1.771 & 3.442 & 3.122 \\
 4 & 1.236 & 3.516 & 3.495 \\   
\hline
\end{tabular}
\end{center}
\end{table}

\end{document}